\pdfoutput=1
\documentclass{JHEP3}
\usepackage{amsmath}

\def\({\left(}
\def\){\right)}
\def\<{\langle}
\def\>{\rangle}
\def\ap{\alpha}

\newcommand{\be}{\begin{equation}}
\newcommand{\ee}{\end{equation}}
\newcommand{\bal}{\begin{aligned}}
\newcommand{\eal}{\end{aligned}}

\newcommand{\labell}[1]{\label{#1}}

\title{Triangulation-free Trivialization of 2-loop MHV Amplituhedron}

\author{Ryota Kojima$^{a}$\footnote{Email: ryota@post.kek.jp}~,
Junjie Rao$^{b}$\footnote{Email: jrao@aei.mpg.de (Corresponding author)}\\
{$^a$KEK Theory Center, Tsukuba, Ibaraki, 305-0801, Japan}\\
{$^b$Max Planck Institute for Gravitational Physics (Albert Einstein Institute), 14476 Potsdam, Germany}}

\abstract{This article introduces a new approach to implement positivity for the 2-loop $n$-particle MHV amplituhedron,
circumventing the conventional triangulation with respect to positive variables of each cell carved out by the sign flips.
This approach is universal for all linear positive conditions and hence free of case-by-case triangulation, as an application
of the trick of positive infinity first introduced in 1910.14612 for the multi-loop 4-particle amplituhedron. Moreover,
the proof of 2-loop $n$-particle MHV amplituhedron in 1812.01822 is revised, and we explain the nontriviality and difficulty
of using conventional triangulation while the results have a simple universal pattern. A further example is presented
to tentatively explore its generalization towards handling multiple positive conditions at 3-loop and higher.}

\preprint{KEK-TH-2241}


\begin{document}
\maketitle

\newpage
\section{Introduction}

The amplituhedron proposal of planar $\mathcal{N}\!=\!4$ SYM \cite{Arkani-Hamed:2013jha,Arkani-Hamed:2013kca} is a
novel reformulation which only uses positivity conditions for all physical and some auxiliary spurious poles to construct
the amplitude or integrand.
For given $(n,k,L)$ where $n$ is the number of external particles, $(k\!+\!2)$ is the number of negative helicities and
$L$ is the loop order, the most generic loop amplituhedron is defined via
\be
Y_\ap^I=C_{\ap a}Z_a^I,~~\mathcal{L}_{(i)\ap}^I=D_{(i)\ap a}Z_a^I,
\ee
here $C_{\ap a}$ is the $(k\!\times\!n)$ positive Grassmannian encoding the tree-level information and $D_{(i)\ap a}$ is the
$(2\!\times\!n)$ positive Grassmannian with respect to the $i$-th loop, and $Z_a^I$ is the kinematical data made of $n$
generalized $(k\!+\!4)$-dimensional momentum twistors, which also obeys positivity as
\be
\<Z_{a_1}\ldots Z_{a_{k+4}}\>>0~~\textrm{for}~~a_1<\ldots<a_{k+4}.
\ee
Based on these notions, we have $\<YZ_iZ_{i+1}Z_jZ_{j+1}\>\!>\!0$, and the rest physical poles involving the
loop part also must be positive:
\be
\<Y\mathcal{L}_{(i)}Z_jZ_{j+1}\>>0,~~\<Y\mathcal{L}_{(i)}\mathcal{L}_{(j)}\>>0.
\ee
To pedagogically investigate this object, we often separately consider the ``pure loop part'' of which $k\!=\!0$, namely
the MHV sector \cite{Arkani-Hamed:2018rsk,Kojima:2018qzz}, in particular the 4-particle case ($n\!=\!4$) has been extensively
understood up to high loop levels \cite{Rao:2017fqc,Rao:2019wyi} and can be compared to the known results from 2-loop to
10-loop level \cite{Bern:1997nh,Bourjaily:2016evz}, and the ``pure tree part'' of which $L\!=\!0$
\cite{Damgaard:2019ztj,Ferro:2020lgp,Ferro:2020ygk}, as well as the simplest nontrivial mixture of these two: the 1-loop
NMHV case ($k\!=\!1, L\!=\!1$) treated in \cite{Kojima:2020tjf}.

Below we will focus on the 2-loop MHV amplituhedron determined by the sign flips \cite{Arkani-Hamed:2017vfh} plus a single
mutual positive condition, and the relevant background of its integrand can be found in
\cite{ArkaniHamed:2010kv,ArkaniHamed:2010gh,Bourjaily:2015jna}. There are also interesting investigations of loop
amplituhedron and the integrated results \cite{Prlina:2017azl,Prlina:2017tvx}.

Back to the specified object of interest, now we continue to elaborate following the general definition above.
As $k\!=\!0$, there is no $Y\!=\!C\!\cdot\!Z$ part, and $L\!=\!2$ gives two $\mathcal{L}\!=\!D\!\cdot\!Z$ parts, each of
which individually obeys physical constraint (note the twisted cyclicity $Z_{n+1}\!=\!-Z_1$ for $k\!=\!0$)
\be
\<\mathcal{L}_{(i)}Z_jZ_{j+1}\>>0,
\ee
and together they obey the mutual positive condition
\be
\<\mathcal{L}_{(1)}\mathcal{L}_{(2)}\>>0.
\ee
First, to triangulate the trivial 1-loop MHV amplituhedron and identify each cell, we need to impose the sign-flip
constraint \cite{Arkani-Hamed:2017vfh}, namely in the sequence (defining $\mathcal{L}_{(1)}\!\equiv\!AB$
and $\mathcal{L}_{(2)}\!\equiv\!CD$ below)
\be
\<AB12\>\!+~~\<AB13\>\!\pm~~\<AB14\>\!\pm~~\ldots~~\<AB1,n\!-\!1\>\!\pm~~\<AB1n\>+,
\ee
while the head $\<AB12\>$ and tail $\<AB1n\>$ are both positive, the entire sequence has two sign flips, so there are
$(n\!-\!2)(n\!-\!3)/2$ possibilities. Explicitly, if the two sign flips occur at
\be
\ldots~~\<AB1i\>\!+~~\<AB1,i\!+\!1\>\!-~~\ldots~~\<AB1j\>\!-~~\<AB1,j\!+\!1\>\!+~~\ldots
\ee
with $2\!\leq\!i\!<\!j\!\leq\!n\!-\!1$, we can parameterize $\mathcal{L}_{(1)}\!=\!AB$ as
\be
A=Z_1+x_1Z_i+w_1Z_{i+1},~B=-Z_1+y_1Z_j+z_1Z_{j+1}, \labell{eq-3}
\ee
which satisfies physical constraint $\<ABZ_aZ_{a+1}\>\!>\!0$ and the sign-filp constraint, similar for
$\mathcal{L}_{(2)}\!=\!CD$:
\be
C=Z_1+x_2Z_k+w_2Z_{k+1},~D=-Z_1+y_2Z_l+z_2Z_{l+1},
\ee
where $x_1,w_1,y_1,z_1$ and $x_2,w_2,y_2,z_2$ are all positive variables. Then, the nontrivial mutual positive condition
of major concern is
\be
\<\mathcal{L}_{(1)}\mathcal{L}_{(2)}\>=\<ABCD\>>0
\ee
for each composite 2-loop cell made of any two 1-loop cells, so there are $(n\!-\!2)^2(n\!-\!3)^2/4$ combinations.

Note that, there are two types of triangulation. The first type is the sign-flip triangulation to carve out each 1-loop
cell with a specific parameterization, while the second is the triangulation with respect to positive variables
of each cell, identical to those extensively manipulated in the 4-particle case which has only one cell.
In this work the triangulation mentioned is the second type, and we will see how the idea of positive infinity
\cite{Rao:2019wyi} can free us from this tedious task first at 2-loop, in an extremely simple way.

\section{Minimal Review of Positive $d\log$ Forms and Dimensionless Ratios}

To get familiar with the mathematical concepts we will extensively use, let's first give a minimal review of $d\log$ forms
in positive geometry. As defined in \cite{Arkani-Hamed:2013kca}, for a positive variable $x$ without further restriction,
we know its $d\log$ form is
\be
\frac{dx}{x}
\ee
which has a singularity at $x\!=\!0$. If we require $x\!>\!a$, as the singularity is shifted to $x\!=\!a$, then the form is
\be
\frac{dx}{x-a},
\ee
on the other hand, the form for $x\!<\!a$ is defined as the complement of $x\!>\!a$ (dropping the measure $dx$):
\be
\frac{1}{x}-\frac{1}{x-a}=\frac{a}{x(a-x)}
\ee
which naturally has two singularities at $x\!=\!0$ and $x\!=\!a$. This equality is also known as the completeness
relation \cite{Rao:2017fqc}, if we reshuffle it as
\be
\frac{1}{x-a}+\frac{a}{x(a-x)}=\frac{1}{x},
\ee
furthermore if we drop the measure $d\log x\!=\!dx/x$ instead of $dx$, we get the completeness relation
\be
\frac{x}{x-a}+\frac{a}{a-x}=1
\ee
in terms of dimensionless ratios \cite{Rao:2019wyi}, which is a more natural way to characterize positive $d\log$ forms.
Here  $x$ and $a$ are treated on the same footing ($a$ also can be a variable), and the sum is always unity.

As done in \cite{Rao:2017fqc}, we can generalize these conditions to $\sum_nx_i\!>\!a$ and $\sum_nx_i\!<\!a$, and the
corresponding dimensionless ratios also sum to unity:
\be
\frac{\sum_nx_i}{\sum_nx_i-a}+\frac{a}{a-\sum_nx_i}=1.
\ee
To inductively prove the dimensionless ratio of $\sum_nx_i\!>\!a$ is
\be
\frac{\sum_nx_i}{\sum_nx_i-a},
\ee
we can first assume it holds for $\sum_{n-1}x_i\!>\!a$. Then depending on $\sum_{n-1}x_i\!\gtrless\!a$, we require
simply $x_n\!>\!0$ or $x_n\!>\!a\!-\!\sum_{n-1}x_i$ to satisfy $\sum_nx_i\!>\!a$, which gives
\be
\frac{\sum_{n-1}x_i}{\sum_{n-1}x_i-a}\times1+\frac{a}{a-\sum_{n-1}x_i}\times\frac{x_n}{x_n-\(a-\sum_{n-1}x_i\)}
=\frac{\sum_nx_i}{\sum_nx_i-a}
\ee
as expected. And $a$ also can be generalized to a sum of positive variables, then the dimensionless ratio of
$\sum_nx_i\!>\!a\!=\!\sum_my_j$ is
\be
\frac{\sum_nx_i}{\sum_nx_i-\sum_my_j}, \labell{eq-1}
\ee
which treats $\sum_nx_i$ and $\sum_my_j$ on the same footing. If any $x_i$ goes to positive infinity \cite{Rao:2019wyi},
the ratio above trivially becomes 1, as any $x_i\!\to\!\infty$ trivializes $\sum_nx_i\!-\!\sum_my_j\!>\!0$.

The physical meaning of positive infinity is also transparent, for example, if we look at
$A\!=\!Z_1\!+\!x_1Z_i\!+\!w_1Z_{i+1}$ in \eqref{eq-3}, while $x_1\!=\!0$ means $A$ is spanned by $Z_1$ and $Z_{i+1}$ only,
$x_1\!\to\!\infty$ leads to $A\!\to\!\infty Z_i$. Recall that a momentum twistor is invariant under rescaling, so this is
equivalent to setting $A\!=\!Z_i$. In fact, using this cut upon positive infinity helps cover configuration $A\!=\!Z_i$
without changing the parameterization into $A\!=\!x_1Z_1\!+\!Z_i\!+\!w_1Z_{i+1}$ and setting $x_1\!=\!w_1\!=\!0$.
As we will see, positive infinity is an indispensable notion for fully understanding the cut structure of the
loop amplituhedron.

Now we are ready to move forward, to explore the extraordinary simplicity hidden in the 2-loop MHV amplituhedron.

\newpage
\section{Triangulation-free Trivialization for Linear Polynomials}

For a single positive condition defining (a cell of) the 2-loop MHV amplituhedron, it's an ubiquitous fact that the
numerator part of its relevant $d\log$ form is always ``maximally positive'', instead of just positive. For example,
in the 4-particle case, if we look at the dimensionless ratio of its $d\log$ form
\be
\frac{D_{12}^+}{D_{12}}=\frac{x_2z_1+x_1z_2+y_2w_1+y_1w_2}{(x_2-x_1)(z_1-z_2)+(y_2-y_1)(w_1-w_2)}
\ee
which is the nontrivial factor in its loop integral (superscript `$+$' is the ``positive monomials extraction'')
\be
\int\frac{dx_1}{x_1}\frac{dy_1}{y_1}\frac{dz_1}{z_1}\frac{dw_1}{w_1}
\frac{dx_2}{x_2}\frac{dy_2}{y_2}\frac{dz_2}{z_2}\frac{dw_2}{w_2}\times\frac{D_{12}^+}{D_{12}},
\ee
obviously $D_{12}^+$ is the maximally positive part of $D_{12}$, namely the term-wise positive polynomial including
every positive term in $D_{12}$. This pattern also applies to all $n\!\geq\!5$ particle cases, and usually the proof must
be done case by case with triangulation. We are often annoyed by the fact that, the tedious triangulation is inevitable but
still this process leaves no trace in the final sum, which however means the sum is correct. This subtle phenomenon motivates
us to circumvent the triangulation, and maybe it is possible to redefine positive conditions that characterize the generic
multi-loop MHV amplituhedron in this way.

First for a single positive condition, so far all cases we have encountered are \textit{linear} in all variables. So we can
assume this polynomial takes a not-so-general form as
\be
P\,(\{x_i\},\{y_j\},\{z_k\})=P_0\,(\{y_j\},\{z_k\})+\sum x_i\,P_i\,(\{y_j\},\{z_k\})>0, \labell{eq-2}
\ee
where $\{x_i\},\{y_j\},\{z_k\}$ are three subsets of all positive variables, and $P_0$ and $P_i$ are independent of
any $x_i$. Such an expansion is always possible, and we can further expand $P_i$ as
\be
P_i\,(\{y_j\},\{z_k\})=P_{i,0}\,(\{z_k\})+\sum y_j\,P_{i,j}\,(\{z_k\})\,.
\ee
Obviously, this nested expansion can be done for as many levels as needed, while this not-so-general form has only three
levels of expansion but it is enough for an inductive proof. Let's give some examples:
\be
P=1-(x+y+z),~~
Q=1-x(1-y(1-z)),~~
R=(1-x)(1-y)(1-z),
\ee
all of these are linear polynomials of level $1,3,3$ respectively. Now assuming the positive sub-condition
\be
P_i=P_{i,0}+\sum y_j\,P_{i,j}>0,
\ee
we want to determine its dimensionless ratio
\be
\frac{P'_i}{P_i}.
\ee
First, $P'_i$ must also be linear in $\{y_j\}$, because a $y_j^2$ term will render this ratio diverge at
$y_j\!=\!\infty$, so will a $1/y_j$ term at $y_j\!=\!0$. Recall that when the positive condition is trivialized, this
ratio must be 1. Therefore we can take the following ansatz
\be
\frac{P'_i}{P_i}=\frac{P'_{i,0}+\sum y_j\,P'_{i,j}}{P_{i,0}+\sum y_j\,P_{i,j}},
\ee
then if all $y_j\!=\!0$, we have a simplified ratio
\be
\frac{P'_i}{P_i}=\frac{P'_{i,0}}{P_{i,0}}=\frac{P^+_{i,0}}{P_{i,0}}.
\ee
Note that $P_{i,0}$ is of level one without further nontrivial entanglement as assumed, for example
\be
P_{i,0}=1-\sum z_k,
\ee
which trivially leads to $P'_{i,0}\!=\!P^+_{i,0}$, as we have proved via \eqref{eq-1} in the previous section.
Next, inspired by the trick of positive infinity in \cite{Rao:2019wyi}, each $y_j\!=\!\infty$ also leads to
$P'_{i,j}\!=\!P^+_{i,j}$. Since $P^+_{i,0},P^+_{i,j}$ are independent of any $y_j$, we must have
\be
\frac{P'_i}{P_i}=\frac{P^+_{i,0}+\sum y_j\,P^+_{i,j}}{P_{i,0}+\sum y_j\,P_{i,j}}=\frac{P^+_i}{P_i},
\ee
again this ``prime'' operation is actually the positive monomials extraction.  Because the derivation above is
inductive, similarly for
\be
P=P_0+\sum x_i\,P_i=P_0+\sum x_i\(P_{i,0}+\sum y_j\,P_{i,j}\)>0,
\ee
we also have
\be
\frac{P'}{P}=\frac{P^+_0+\sum x_i\,P^+_i}{P_0+\sum x_i\,P_i}=\frac{P^+}{P},
\ee
which finishes the clean proof of the dimensionless ratio for a linear $P$ of any levels of nested expansion.

The physical meaning of this proof is, given a generic ansatz $P^+/P$ based on extensive known results we find that
it is the only legal quantity that satisfies the correct cut structure simplified by cuts at either zero or infinity,
and here positivity is the only dominating principle. As we have explained in the previous section, cuts upon positive
infinity are equivalent to conventional cuts upon zero after reparameterization.

\section{2-loop MHV Amplituhedron Revisited}

Now for a generic cell of the 2-loop MHV amplituhedron \cite{Kojima:2018qzz,Arkani-Hamed:2017vfh}, from the parameterization
\be
\bal
A=Z_1+x_1Z_i+w_1Z_{i+1},~B=-Z_1+y_1Z_j+z_1Z_{j+1},\\
C=Z_1+x_2Z_k+w_2Z_{k+1},~D=-Z_1+y_2Z_l+z_2Z_{l+1},
\eal
\ee
we see the mutual positive condition
\be
\<ABCD\>=\<Z_1\!+\!x_1Z_i\!+\!w_1Z_{i+1},-Z_1\!+\!y_1Z_j\!+\!z_1Z_{j+1},Z_1\!+\!x_2Z_k\!+\!w_2Z_{k+1},
-Z_1\!+\!y_2Z_l\!+\!z_2Z_{l+1}\>>0
\ee
is a linear polynomial, and it can have maximally four levels. Let's see a concrete example by choosing
\be
i=2,~j=8,~k=4,~l=6,
\ee
then this quantity becomes
\be
\bal
\<ABCD\>=\,\,&\<Z_1\!+\!x_1Z_2\!+\!w_1Z_3,Z_1\!+\!x_2Z_4\!+\!w_2Z_5,-Z_1\!+\!y_2Z_6\!+\!z_2Z_7,-Z_1\!+\!y_1Z_8\!+\!z_1Z_9\>,\\
=\,\,&C+x_1(-\,C_2+x_2(-\,C_{2,4}+y_2C_{2,4,6}+z_2C_{2,4,7})+w_2(-\,C_{2,5}+y_2C_{2,5,6}+z_2C_{2,5,7}))\\
&~~\,+w_1(-\,C_3+x_2(-\,C_{3,4}+y_2C_{3,4,6}+z_2C_{3,4,7})+w_2(-\,C_{3,5}+y_2C_{3,5,6}+z_2C_{3,5,7}))\,,
\eal
\ee
where the positive determinants are defined as
\be
\bal
C=\,&\<Z_1,x_2Z_4\!+\!w_2Z_5,y_2Z_6\!+\!z_2Z_7,y_1Z_8\!+\!z_1Z_9\>,\\
C_i=\,&\<Z_1,Z_i,y_2Z_6\!+\!z_2Z_7,y_1Z_8\!+\!z_1Z_9\>,\\
C_{i,j}=\,&\<Z_1,Z_i,Z_j,y_1Z_8\!+\!z_1Z_9\>,\\
C_{i,j,k}=\,&\<Z_i,Z_j,Z_k,-Z_1+y_1Z_8\!+\!z_1Z_9\>,
\eal
\ee
we see that it actually has three levels, since $C_{i,j,k}$ is trivially positive and needs not expand as a fourth.
Then we can immediately apply the proof in the previous section to show that its dimensionless ratio is
\be
\frac{\<ABCD\>^+}{\<ABCD\>},
\ee
note that if we try to prove this result with triangulation, it will be extremely tedious already for the 2-loop case,
as we have to handle complicated shifting and intersecting relations in three copies of 2-dimensional planes spanned by
variables $(x_1,w_1)$, $(x_2,w_2)$ and $(y_2,z_2)$. Since such a proof holds for generic $i,j,k,l$, all $d\log$ forms
corresponding to various cells of the 2-loop MHV amplituhedron are trivialized and free of the case-by-case triangulation.

\section{Proof in 1812.01822 Revised}

However, the proof in \cite{Kojima:2018qzz} using triangulation also seems clean (see its Appendix B). Now we explain
why this proof should be revised while its conclusion still holds. There, the positive condition of combination
$i\!<\!k\!<\!l\!<\!j$ is reorganized as (the arguments indicate how $a,b,c,d,e$ depend on $z_2,w_1,x_1,w_2,y_2$)
\be
\<ABCD\>=a(w_1,x_1,w_2)\,z_2-b(z_2,y_2)\,w_1-c(z_2,w_2,y_2)\,x_1-d(w_1)\,w_2+e(w_1,x_1,w_2)\,y_2,
\ee
then the subsequent discussion continues as if these $a,b,c,d,e$ were all constants. We find it problematic,
because this is equivalent to rescaling $z_2,w_1,x_1,w_2,y_2$ by five constants respectively, but the Jacobian of this
rescaling is not trivially 1. So why is the conclusion still correct?

The subtle secret here is that though the rescaling is illegal, $a,b,c,d,e$ are still positive. So pretending that
they were constants just gives us the same result
\be
\frac{a\,z_2+e\,y_2}{a\,z_2-b\,w_1-c\,x_1-d\,w_2+e\,y_2}=\frac{\<ABCD\>^+}{\<ABCD\>},
\ee
while the correct logic is not so trivial. Without the trick of positive infinity, we will have a tough work of
triangulation to do. Now we find an even cleaner and also more general proof for this neat result.

\section{An Example of Quasi-linear Polynomials}

Besides linear polynomials, we would like to go further and take a glance at an interesting generalization:
an example of the \textit{quasi-linear} polynomials, which is a linear polynomial times an overall positive factor,
though we will not explore this category systematically as we have done before.

In this case, the problem originates from a 3-loop example proved in \cite{Rao:2017fqc} (namely $T_8$) which
has three positive conditions:
\be
z_1+c_{12}>z_2,~~z_1+c_{13}>z_3,~~z_2+c_{23}>z_3,
\ee
besides $z_1,z_2,z_3$, here $c_{12},c_{13},c_{23}$ are also treated as independent positive variables
(namely intermediate variables introduced in \cite{Rao:2017fqc}). Using ordinary triangulation, we have known its
dimensionless ratio is
\be
\frac{(z_1+c_{12})(z_1+c_{13})(z_2+c_{23})-z_1z_2z_3}{(z_1+c_{12}-z_2)(z_1+c_{13}-z_3)(z_2+c_{23}-z_3)},
\ee
now let's see how the new proof reproduces this result.

First, since the new proof can only handle a single positive condition, we have to trivialize two out of three by
defining some convenient positive variables as below:
\be
z_2\equiv\frac{s}{1+s}\,(z_1+c_{12}),~~z_3\equiv\frac{t}{1+t}\,(z_1+c_{13}),
\ee
obviously, when $s$ ranges from 0 to $\infty$, $z_2$ naturally ranges from 0 to $(z_1\!+\!c_{12})$,
and similar for $t$ and $z_3$. It is easy to find their reverse transformations, given by
\be
s=\frac{z_2}{z_1+c_{12}-z_2},~~t=\frac{z_3}{z_1+c_{13}-z_3},
\ee
and clearly this change of variables is non-linear. In terms of $s,t$, the third condition becomes
\be
\frac{s}{1+s}\,(z_1+c_{12})+c_{23}-\frac{t}{1+t}\,(z_1+c_{13})>0,
\ee
or equivalently
\be
\frac{c_{23}+(c_{12}+c_{23}+z_1)\,s+((c_{23}-c_{13}-z_1)+(c_{12}+c_{23}-c_{13})\,s)\,t}{(1+s)(1+t)}>0.
\ee
Note the numerator is of the form $(A(s)\!+\!B(s)t)$ as $A,B$ do not depend on $t$, which is linear in all variables
and separated properly as \eqref{eq-2}. Now forgetting the positive denominator, we can safely use the new proof for this
quasi-linear polynomial and obtain the dimensionless ratio
\be
R_8=\frac{c_{23}+(c_{12}+c_{23}+z_1)\,s+(c_{23}+(c_{12}+c_{23})\,s)\,t}
{c_{23}+(c_{12}+c_{23}+z_1)\,s+((c_{23}-c_{13}-z_1)+(c_{12}+c_{23}-c_{13})\,s)\,t},
\ee
multiplied by the Jacobian transformed from $(s,t)$ back to $(z_2,z_3)$ and factors from $d\log$ forms, it is then
\be
\frac{\partial(s,t)}{\partial(z_2,z_3)}\frac{z_2\,z_3}{s\,t}\,R_8
=\frac{(z_1+c_{12})(z_1+c_{13})(z_2+c_{23})-z_1z_2z_3}{(z_1+c_{12}-z_2)(z_1+c_{13}-z_3)(z_2+c_{23}-z_3)}
\ee
as expected. Furthermore, in \cite{Rao:2017fqc} there are other seven $d\log$ forms (namely $T_1\ldots T_7$) that can
be obtained by flipping $c_{ij}$ to $-c_{ji}$ in the denominator and setting $c_{ij}$ to zero in the numerator, which
exactly reflects the logic of the new proof as the numerator always collects positive terms only.

This example also provides a tentative approach to extend the triangulation-free trivialization to the cases with multiple
positive conditions, as will be discussed more in the next section. But of course, we should note this example is a much
simpler case in the context of 4-particle amplituhedron, as restricted to the ordered subspace $X(123)$ in which
$x_1\!<\!x_2\!<\!x_3$. In general, $n\!\geq\!5$ particle cases at 3-loop will have various combinations of 1-loop cells
in terms of three sets of loop variables, so the positive conditions are no longer uniform, and they may have more
complicated nested expansions.

\section{Outlook}

The discussion above is also a key motivation to develop a triangulation-free approach, otherwise even
the 3-loop work will be overwhelmingly difficult. The luxurious ambition is to extend the 2-loop proof to the
all-loop, generic $n$-particle MHV amplituhedron, or directly redefine this geometric object with positivity
but without the annoying triangulation. Here, we can easily trivialize positivity by evaluating the integral at zero or
positive infinity with respect to some variables, however, unlike the 2-loop case, its challenge is to reconstruct the
correct integrand or dimensionless ratio from multiple positive conditions. How to find a minimal set of such ``cuts''
that can fully cover every facets of the object, requires a further geometric understanding, especially about the
shifting and intersecting relations among multiple higher dimensional planes representing the positive constraints.

Naturally, the 4-particle amplituhedron at 3-loop is a simplest nontrivial testing ground for this goal of which the result
has been well known from various perspectives, and more importantly, in the 4-particle case, the positive conditions are
always uniform and this symmetry is partly maintained upon the cuts. In fact, the Mondrian reduction \cite{Rao:2019wyi}
is a special type of application of these cuts, but we must know the DCI integral basis first in that diagrammatic context,
and now we would like to derive the basis as well from a more algebraic perspective, as for the generic $n\!\geq\!5$
particle case there is no simple insight similar to the Mondrian diagrammatics. In the future, we will focus on the
4-particle case up to higher loops as usual, as well as the tentative derivation of the 5-particle case at 3-loop, using
the triangulation-free approach.

\section*{Acknowledgments}

The authors would like to thank Nima Arkani-Hamed for giving valuable comments on the manuscript.

\newpage

\end{document}